\documentclass[]{spie}  

\usepackage{amsmath,amsfonts,amssymb}
\usepackage{graphicx}
\usepackage[colorlinks=true, allcolors=blue]{hyperref}
\usepackage{multirow}
\newcommand{\bftab}{\fontseries{b}\fontsize{10pt}{10pt}\selectfont}
\usepackage{xcolor}

\title{PRISM Lite: A lightweight model for interactive 3D placenta segmentation in ultrasound}

\author[a]{Hao Li}
\author[b]{Baris Oguz}
\author[b]{Gabriel Arenas}
\author[a]{Xing Yao}
\author[a]{Jiacheng Wang}

\author[b]{Alison Pouch}
\author[a]{\\Brett Byram}
\author[b]{Nadav Schwartz}
\author[a]{Ipek Oguz}

\affil[ ]{\textsuperscript{a}Vanderbilt University, \textsuperscript{b}University of Pennsylvania}

\begin{document} 
\maketitle

\begin{abstract}

Placenta volume measured from 3D ultrasound (3DUS) images is an important tool for tracking the growth trajectory and is associated with pregnancy outcomes. Manual segmentation is the gold standard, but it is time-consuming and subjective. Although fully automated deep learning algorithms perform well, they do not always yield high-quality results for each case. Interactive segmentation models could address this issue. However, there is limited work on interactive segmentation models for the placenta. Despite their segmentation accuracy, these methods may not be feasible for clinical use as they require relatively large computational power which may be especially prohibitive in low-resource environments, or on mobile devices. In this paper, we propose a lightweight interactive segmentation model aiming for clinical use to interactively segment the placenta from 3DUS images in real-time. The proposed model adopts the segmentation from our fully automated model for initialization and is designed in a human-in-the-loop manner to achieve iterative improvements. The Dice score and normalized surface Dice are used as evaluation metrics. The results show that our model can achieve superior performance in segmentation compared to state-of-the-art models while using significantly fewer parameters. Additionally, the proposed model is much faster for inference and robust to poor initial masks. The code is available at \url{https://github.com/MedICL-VU/PRISM-placenta}.

\end{abstract}


\section{INTRODUCTION}
Placental size and shape are associated with adverse perinatal outcomes \cite{biswas2008gross,redman2005latest,schwartz2014first} and fetal size \cite{schwartz2012two}. Volume measurement using 3D ultrasound images (3DUS) is crucial for assessing potential perinatal morbidity and mortality. Although human annotation is the gold standard for measuring the volume, it is time consuming and subjective. Recently, deep learning-based automated methods \cite{liu2023medical} have shown state-of-the-art performance for placenta segmentation \cite{looney2018fully,looney2021fully,looney2017automatic,oguz2018combining,pouch2020automated,schwartz2022fully,zimmer2023placenta}. However, due to the poor image quality, noise, and artifacts in 3DUS, the standard deviation ranges of these methods indicate they may not consistently produce robust segmentation in challenging cases. Interactive segmentation methods may overcome this by leveraging user input to specify the target region.

Recently, the Segment Anything Model (SAM) \cite{Kirillov_2023_ICCV} has shown superior performance and wide generalizability for segmentation tasks by taking visual prompts, such as points and boxes, and it has been widely adopted for medical image segmentation tasks \cite{ma2024segment,li2023promise,li2023assessing,gong20233dsam,wang2023novel,wang2023sam,cheng2023sam,li2024prism}, including for US applications \cite{yao2023false,lin2023samus,tu2024ultrasound,li2024inter}, to produce robust segmentation. 
Among these SAM-based methods, the PRISM \cite{li2024prism,li2024inter} stands out due to its ability to produce segmentations that can reach the human level. It is designed in a human-in-the-loop manner to allow iterative corrections to achieve substantial improvements. It takes points, boxes, scribbles, and masks as prompts for robust outcomes. Its effectiveness and efficiency have been proven  in the context of placenta segmentation from 3DUS \cite{li2024inter}.


Nonetheless, a limitation of PRISM is that even though it can accept masks produced from the previous iterations as prompt input to achieve continual improvements, PRISM starts interactive segmentation from scratch (i.e., requiring substantial user initialization), instead of adapting a binary mask produced from a pretrained model. This contrasts with more practical minimally interactive strategies where directly adapting a mask from a pretrained model would be preferable in order to more efficiently meet the user expectations. 


PRISM has two limitations that, if overcome, could lead to clinical adoption.
First, it uses a hybrid network \cite{li2022cats} as an image encoder, which requires substantial computational resources. Second, the large model size slows down the computation. These limit its usability in scenarios such as implementation on mobile devices or directly on the ultrasound console, or for challenging images that require many iterative corrections. This also lowers its usability in low-resource environments where point-of-care US (POCUS) may be available.

In this paper, we propose a lightweight interactive segmentation model with low computational needs. The proposed model adapts an initial mask from an automated model \cite{schwartz2022fully}, and we design it to be compact to meet the requirements of resource-limited environments. The human-in-the-loop strategy is applied for iterative improvements. The evaluation is conducted on 3DUS images for placenta segmentation. Our results indicate that the proposed framework is more effective, efficient, and robust to poor initial masks. It also requires less computational resources and has a faster inference speed.

\section{Materials and Methods}

\noindent\underline{\textbf{Dataset.}}
3D ultrasound volume datasets were used in this study that were acquired from pregnant women (n=124) at 11–14 weeks of gestation using GE Voluson E8 ultrasound machines. The dimensions of the raw images ranged from $245 \times 265 \times 173$ to $714 \times 726 \times 488$ voxels, with a mean isotropic resolution of $0.49 \pm 0.04 \mathrm{mm}$.  The images are resampled to a $1$mm isotropic resolution in the experiments.  We adopted the same data split as a prior study \cite{li2024inter} with a training/validation/testing split of 0.6/0.2/0.2. The test set consisted of 14 anterior and 10 posterior placentas. The details of preprocessing steps and data augmentations can be found in \cite{li2024inter}.


\begin{figure}[t]
\centering
\includegraphics[width=\linewidth]{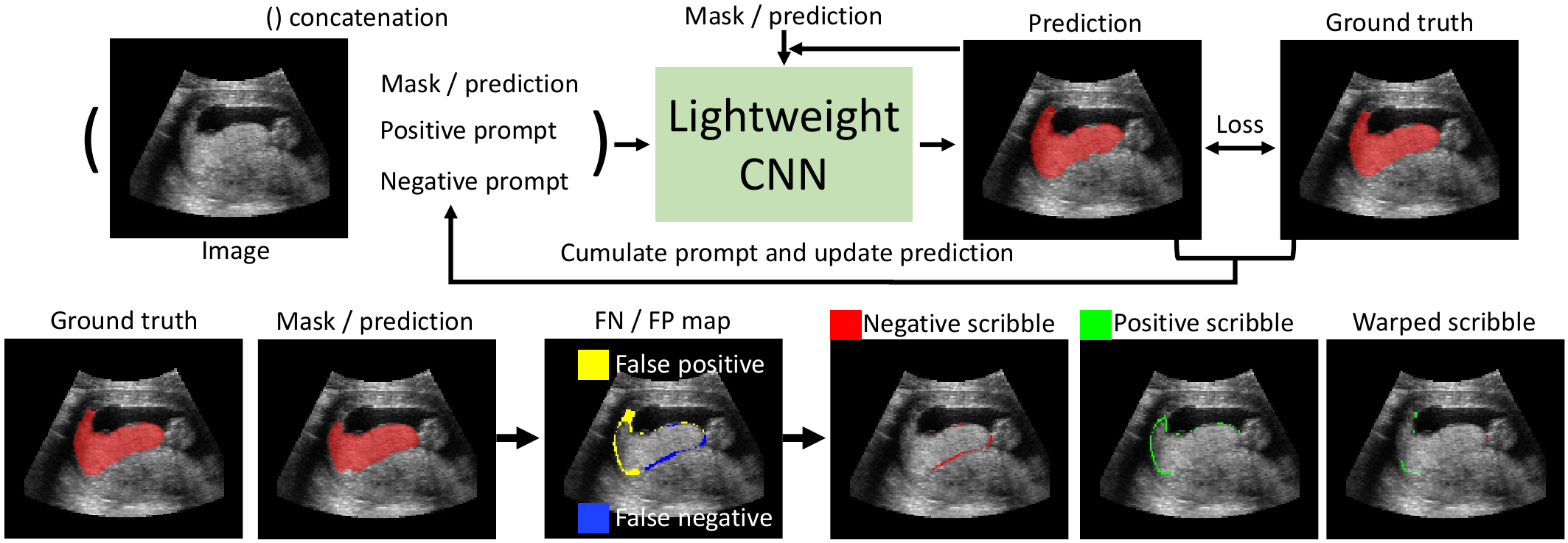}   
\caption{\textbf{Top}, the proposed 3D interactive segmentation framework, illustrated in 2D. \textbf{Bottom}, warped scribble generation. Briefly, they are generated by breaking non-warped scribbles into line segments and warping \cite{li2024inter}.}
\label{framework}
\end{figure}

\noindent\underline{\textbf{Framework.}}
The proposed interactive framework is shown in Fig.~\ref{framework} (top). The model is designed in a human-in-the-loop manner to achieve iterative improvements. Specifically, the proposed framework takes the concatenation of image, mask, and positive/negative prompt maps to produce robust placenta segmentation. The mask is obtained using the automated segmentation model \cite{schwartz2022fully} at the initial iteration. In the subsequent iterations, it is replaced by the prediction from the last iteration. 
We adapt the iterative learning method from previous work \cite{li2024prism} by adding the loss from each iteration and propagating this added loss back to update the model. In this way, each subsequent iteration incorporates information from the last iteration to achieve iterative correction. We consistently set the total iteration number to 11 for training and inference across all subjects. However, in practice, the number of iterations may vary, as users stop once the segmentation results meet their expectations.

To achieve robust segmentation performance with iterative improvements, the framework takes various sparse visual prompts to form cumulative prompt maps to indicate the target region. In the experiments we automatically generate these sparse prompts to mimic human behavior.


\begin{itemize}
\setlength\itemsep{-0.3em}
\item \underline{{Box}}: The 3D bounding box is defined only at the initial iteration based on the ground truth.

\item \underline{{Point}}: At each iteration, positive and negative point prompts are randomly sampled with uniform distribution from the false negative (FN) and false positive (FP) regions, respectively. 

\item \underline{{Warped scribble}}: Scribble generation follows the previous work \cite{wong2023scribbleprompt,li2024inter}. Briefly, we extract the skeletons of the FN and FP regions from mask or prediction (Fig.~\ref{framework}, bottom). A random mask is generated and used to break up the skeletons, and a random deformation field is applied to warp the broken skeletons. We generated the centerline scribbles based on the transverse plane due to the empirically superior  performance \cite{li2024inter}.
\end{itemize}

\begin{figure}[t]
\centering
\includegraphics[width=\linewidth]{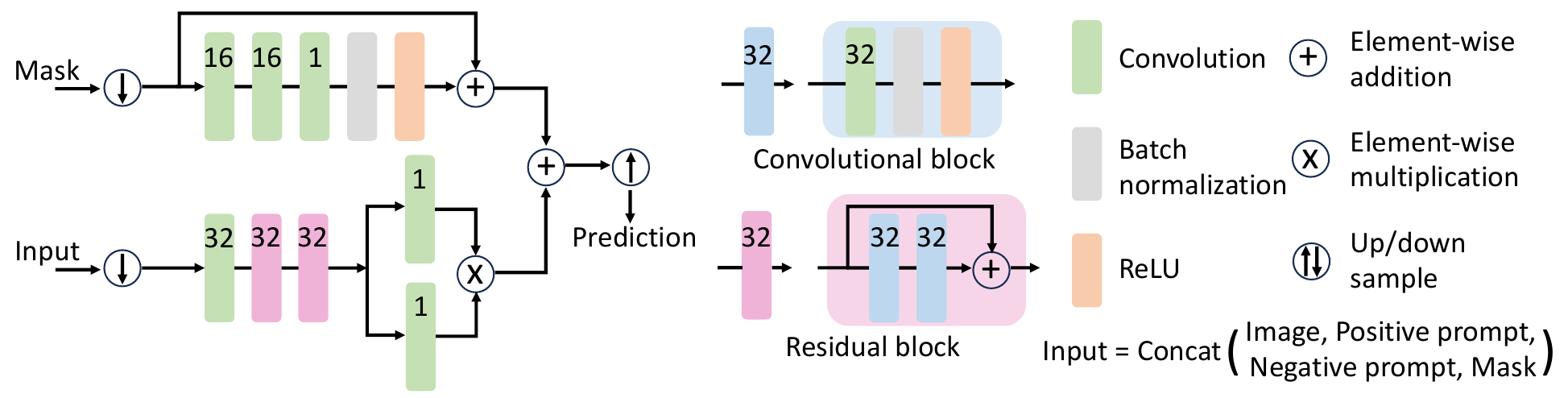}   
\caption{Network architecture of the proposed lightweight model. The numbers of output channels are marked.}
\label{network}
\end{figure}

\noindent\underline{\textbf{Network architecture.}}
Fig.~\ref{network} shows the proposed lightweight 3D segmentation model, designed for scenarios with limited computational resources to ensure robust performance. It has two paths: the top path encodes the binary mask for the initial iteration or the prediction logits map from the previous iteration, while the bottom path takes the concatenation of an image, a mask, and prompt maps as input to ensure segmentation accuracy by incorporating more information. To reduce the parameter numbers, the kernel size of convolutions is set to 3 in residual blocks, and 1 for others. Both the input and mask are downsampled by a factor of 2 to increase processing efficiency. The feature maps from both paths are fused through simple element-wise addition, and the fused information is upsampled to its original resolution for the final prediction. The entire network is compact, and the parameter size is approximately equivalent to the residual block of a 3D network (input channel=32, parameters=0.08M)\cite{li2021longitudinal}. This lightweight model can achieve iterative corrections on the automated segmentation results and produce a superior segmentation that approaches human-level performance.


\noindent\underline{\textbf{Implementation details.}}  The 3D input image size is $128\times128\times128$.
We used the combination of Dice and cross-entropy loss and followed the details in previous work \cite{li2024prism,li2024inter}.  Our study was conducted using an NVIDIA A6000 GPU, an Intel(R) Xeon(R) Silver 4208 CPU @ 2.10GHz, and Ubuntu 22.04.3 LTS as the operating system. The Dice score and normalized surface Dice (NSD) with 1mm as tolerance are used as evaluation metrics. The compared methods include state-of-the-art interactive methods, such as SAM \cite{Kirillov_2023_ICCV}, model adaptation methods \cite{li2023promise,gong20233dsam} from SAM, and iterative methods, as well as a fully automated placenta segmentation method \cite{schwartz2022fully}.

\begin{table*}[b]
 \caption{$Dice / NSD$ results comparison, presented as ${mean} ({std.\ dev}) (\%)$. * uses 1 point, 1 box, and 2D warped centerline scribbles per volume. $\bullet$ uses 1 point per slice. $\dagger$  uses 10 points per volume. Bold indicates best performance. }
 

\label{main_table}
\small
\begin{center}
    \begin{tabular}{ l | c |c |c }
    \hline
    \multicolumn{1}{l|}{Methods} &   \multicolumn{1}{c|}{Anterior} & \multicolumn{1}{c|}{Posterior} & \multicolumn{1}{c}{Overall}\\
    \hline

    \ Automated \cite{schwartz2022fully} &    90.46 (2.89) / 80.72 (7.92) & 89.42 (2.57) / 75.75 (9.21) & 90.03 (2.81) / 78.65 (8.83) \\

    \hline
    $\bullet$SAM \cite{Kirillov_2023_ICCV} & 46.44 (8.93) / 14.29 (4.56)  & 43.32 (8.85) / 14.78 (4.44) &  45.14 (9.02) / 14.49 (4.52) \\

   $\dagger$3DSAM-adapter \cite{gong20233dsam}  & 84.57 (9.89) / 70.34 (14.5)  & 85.97 (5.39) / 68.58 (12.0) & 85.15 (8.34) / 69.61 (13.5) \\

       $\dagger$ProMISe \cite{li2023promise}  &  85.55 (7.74) / 71.57 (11.3)  & 83.91 (6.18) / 67.15 (12.2) & 84.87 (7.17) / 69.73 (11.9)  \\

     $\dagger$SAM-Med3D-turbo \cite{wang2023sam} &  89.51 (2.61) / 76.22 (8.77)  &  88.59 (2.37) / 73.32 (7.51) & 89.13 (2.55) / 75.01 (8.39) \\


     \cline{1-4}

      *PRISM \cite{li2024inter}  &  97.35 \bftab{(.410)} / \normalfont{99.68} \bftab{(.179)}  &  {97.01 (.545) / 99.44 (.236)}  & {97.15 (.520) / 99.54 (.244)} \\

     *PRISM Lite  &  \bftab{98.00}  \normalfont{(.535)} / \bftab{99.84} \normalfont{(.190)}  &  \bftab{98.05 (.485) / 99.87 (.193)}  & \bftab{98.00 (.517)/ 99.86 (.192)} \\
    \hline

\multicolumn{4}{l}{
\begin{tabular}{@{}l@{}@{}} 
    
\end{tabular}
        }  
        
\end{tabular} 

\end{center}
\end{table*}

\begin{table*}[t]
\caption{Model complexity and computational results. FLOP input size is $128\times128\times128$. CPU time is inference time of the model only, and does not include file I/O, post-processing, etc.} 

\label{efficient_table}
\small
\begin{center}
    \begin{tabular}{ l  | c |c |c }
    \hline
    \multicolumn{1}{l|}{Methods} &   \multicolumn{1}{c|}{$\#$Params (M)} & \multicolumn{1}{c|}{GFLOPs} & \multicolumn{1}{c}{CPU time (s)} \\
    \hline


      PRISM \cite{li2024inter} & 118  & 557.82 & 10.46  \\

     PRISM Lite & 0.1  &  29.29 &  0.75 \\
    \hline     


\end{tabular} 

\end{center}
\end{table*}

\section{Results}

Tab.~\ref{main_table} presents a quantitative comparison with state-of-the-art segmentation methods. The automated method outperforms most interactive segmentation methods. However, both PRISM and PRISM Lite demonstrate exceptional performance, achieving accuracy with Dice scores above 0.95. We note that previous reports of inter-rater variability are in the 0.85-0.9 range for manual segmentation \cite{zimmer2023placenta}. PRISM Lite has the best performance among the compared methods, except a slightly larger standard deviation than PRISM for anterior placentas.

Tab.~\ref{efficient_table} compares the model complexity and computational cost, indicating that PRISM Lite is more efficient than PRISM, as expected. PRISM Lite has about 1,000 times fewer trainable parameters than PRISM, which is achieved by adopting the mask from the automated model rather than starting from scratch. The compact model also results in a much lower total number of floating point operations and faster CPU inference time. Importantly, we note that PRISM Lite has better segmentation performance than PRISM (Tab.~\ref{main_table}) despite such efficient use of computational resources (Tab.~\ref{efficient_table}).

\begin{figure}[t]
\centering
\includegraphics[width=\linewidth]{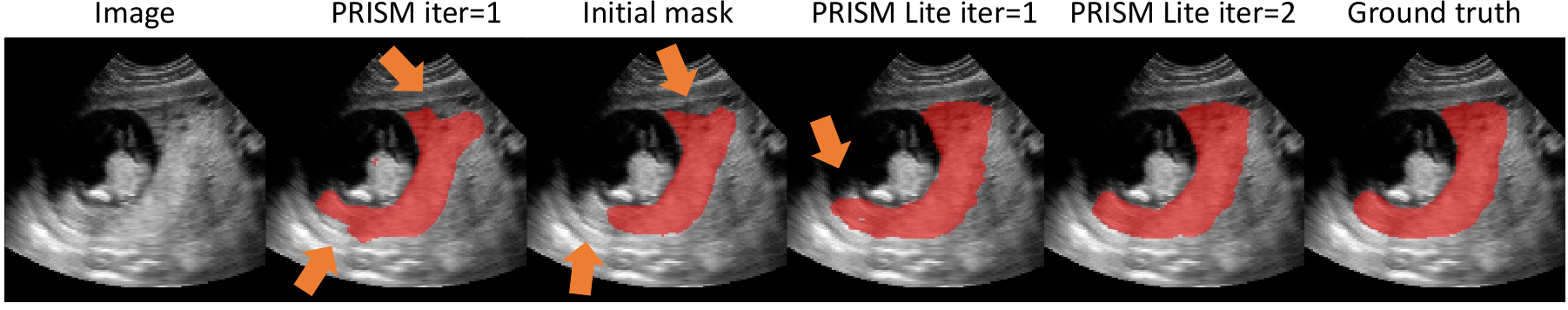}   
\caption{Qualitative results. Orange arrows highlight segmentation errors. PRISM Lite rapidly corrects these errors.}
\label{qualitative}
\end{figure}
\begin{figure}[b]
\centering
\includegraphics[width=\linewidth]{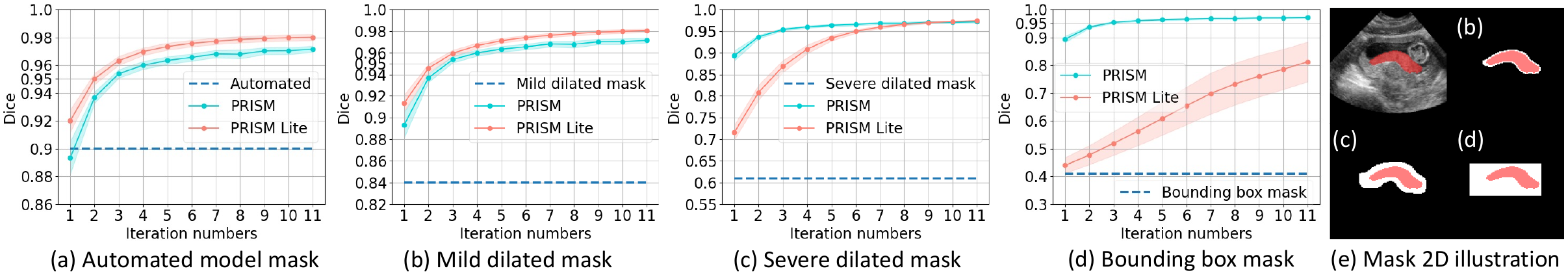}   
\caption{(a) Performance using the automated model result as initial mask for PRISM Lite. (b-d) Performance using polluted initial masks. (e) illustrates these polluted masks in white, overlaid with the raw automated model mask in red.} 
\label{iterative}
\end{figure}

The qualitative results (Fig.~\ref{qualitative}) show that: (1) PRISM Lite produces better results than PRISM at the first iteration, leveraging the mask from the automated model, and (2) PRISM Lite can closely approximate the ground truth within just a few iterations. These are further evidenced in Fig.~\ref{iterative}(a), where PRISM Lite improves the Dice score of the automated segmentation after a single iteration, and outperforms PRISM at every iteration.

Finally, to test the robustness of PRISM Lite to poor quality initial masks, we used three alterations to pollute the automated segmentation result: a mild 3D dilation (kernel of 3 $\mathrm{mm}$), a severe 3D dilation (kernel of 9 $\mathrm{mm}$), and a 3D bounding box, as shown in Fig.~\ref{iterative}(e).  Fig.~\ref{iterative}(b)-(d) show the performance under these pollution conditions. It is easy to see that PRISM Lite can achieve substantial iterative improvements over the input in each condition. However, for masks with significant errors (Fig.~\ref{iterative}(c)-(d)), PRISM is more efficient.

\section{Conclusion}
In this paper, we propose a lightweight model, PRISM Lite, for interactive segmentation of the placenta from 3D ultrasound images. By adopting the result from an automated model as an initial mask prompt, PRISM Lite outperforms state-of-the-art methods. We designed PRISM Lite to be much more compact, requiring significantly lower computational resources and with a faster inference time, which might make it suitable for real-time applications with limited computational resources, such as mobile devices. Importantly, this more compact architecture allows the model to learn more efficient representations and the accuracy of the model is better than the larger PRISM model. 


\clearpage

\acknowledgments{This work was supported, in part, by NIH grants R01-HD109739, R01-HL156034, U01-HD087180, and R03-HD069742, as well as by the Penn Presbyterian George L.\ and Emily McMichael Harrison Fund for Research in Obstetrics and Gynecology.

\bibliography{report} 
\bibliographystyle{spiebib} 

\end{document}